\documentclass[12pt]{iopart}
\usepackage{graphicx}
\usepackage{subfigure}
\usepackage{epsfig}
\usepackage{epstopdf}

\usepackage{iopams}

\begin{document}

\title[Speeding up adiabatic passage with an optimal modified Roland-Cerf protocol]{Speeding up adiabatic passage with an optimal modified Roland-Cerf protocol}

\author{Dionisis Stefanatos$^*$ and Emmanuel Paspalakis}

\address{Department of Materials Science, University of Patras, Patras 265 04, Greece}
\ead{$^*$dionisis@post.harvard.edu}
\vspace{10pt}
\begin{indented}
\item[]June 2019
\end{indented}

\begin{abstract}

In this article we propose a novel method to accelerate adiabatic passage in a two-level system with only longitudinal field (detuning) control, while the transverse field is kept constant. The suggested method is a modification of the Roland-Cerf protocol, during which the parameter quantifying local adiabaticity is held constant. Here, we show that with a simple ``on-off" modulation of this local adiabaticity parameter, a perfect adiabatic passage can be obtained for every duration larger than the lower bound $\pi/\Omega$, where $\Omega$ is the constant transverse field. For a fixed maximum amplitude of the local adiabaticity parameter, the timings of the ``on-off" pulse-sequence which achieves perfect fidelity in minimum time are obtained using optimal control theory. The corresponding detuning control is continuous and monotonic, a significant advantage compared to the detuning variation at the quantum speed limit which includes non-monotonic jumps. The proposed methodology can be applied in several important core tasks in quantum computing, for example to the design of a high fidelity controlled-phase gate, which can be mapped to the adiabatic quantum control of such a qubit. Additionally, it is expected to find applications across all Physics disciplines which exploit the adiabatic control of such a two-level system.

\end{abstract}

\vspace{2pc}
\section{Introduction}

\label{intro}

Controlling efficiently the fundamental quantum unit, the two-level quantum system, lies at the heart of many modern quantum technology applications \cite{Roadmap17,Glaser15}. One of the most effective methods to address this problem is adiabatic passage (AP) \cite{Vitanov01,Goswami03}. The system starts from an eigenstate of the initial Hamiltonian, then some parameter varies slowly with time and, if the change is slow enough, it ends up to an eigenstate of the final Hamiltonian. The traditional setup for AP is a two-level system  where only the longitudinal $z$-field is time-dependent, while the transverse $x$-field is constant. This framework not only describes the setting of some classical applications, for example nuclear magnetic resonance, but is also pertinent to some modern applications, like several important core tasks in quantum computing \cite{Martinis14a,Martinis14,Shim16,Zeng18NJP,Zeng18PRA,Fischer19}. As a concrete example we mention the design of a high fidelity controlled-phase gate \cite{Martinis14a}, which can be mapped to the adiabatic quantum control of such a qubit \cite{Martinis14}.


In the traditional AP, the slow change in the control parameter, $z$-field, is linear, and the process is called Landau-Zener (LZ) sweep \cite{Landau32,Zener32}. The method has been proven to be robust to moderate variations of the system parameters. Its major limitation is, as with every adiabatic method, the necessary long operation time which may lead to a degraded performance in the presence of decoherence and dissipation. In order to speed up the evolution, several methods have been suggested. For example, it has been shown that certain nonlinear LZ sweeps can achieve perfect fidelity for specific durations \cite{Garanin02}. In a related work \cite{Martinis14}, the error probability of the final state with respect to the adiabatic evolution is minimized for durations larger than a certain threshold. A high fidelity is achieved, at levels appropriate for fault-tolerant quantum computation, even for durations as short as a few times the system timescale. Optimal control theory has also been exploited to find the quantum speed limit for the desired transfer \cite{Calarco09,Hegerfeldt13}, but it requires infinite values of the control field in order to implement instantaneous rotations around $z$-axis. More realistic speed limits have been obtained for bounded control \cite{Hegerfeldt14}, but their implementation also requires discontinuous and non-monotonic changes of the $z$-field. Finally, we mention the methods developed under the umbrella of \emph{Shortcuts to Adiabaticity} \cite{Odelin19,Demirplak03,Berry09,Motzoi09,Chen10a,Masuda10,Deffner14}, where the quantum system is driven at the same final state as with a slow adiabatic process, but without necessarily following the instantaneous adiabatic eigenstates at intermediate times. The common characteristic of these techniques when applied to two-level quantum systems \cite{Chen10,Chen11,Bason12,Malossi13,Ruschhaupt12,Daems13,Ibanez13,Motzoi13,Theis18}, is that both the longitudinal ($z$) and transverse ($x$) fields are exploited in order to speed up adiabatic evolution, while here we focus on the restricitve framework where only the $z$-field is time-dependent.

The Roland-Cerf (RC) protocol was originally developed in order to accelerate quantum search in adiabatic quantum computation \cite{Roland02}. It relies on the fulfilment of a local (in time) adiabaticity condition, instead of a global one valid during the whole process. In the present work, we first apply the RC protocol with only detuning ($z$-field) control, as in Refs. \cite{Bason12,Malossi13}, and show that it can achieve perfect fidelity for specific durations, as the nonlinear LZ sweeps. During the application of this protocol, the parameter quantifying local adiabaticity is held constant. Next, we suggest a modified RC protocol, with ``on-off" modulation of the local adiabaticity parameter, which can achieve perfect fidelity for every duration larger than a lower bound. Compared to our recent related work \cite{Stefanatos19}, here we use optimal control theory to obtain an extra optimality condition, see Sec. \ref{sec:optimal_solution}, which allows to determine the timings of the ``on-off" optimal control by solving a single transcendental equation. This is a significant improvement compared to Ref. \cite{Stefanatos19}, where the optimal timings are obtained through a numerical optimization with respect to the control amplitude. The suggested method exploits the advantages of composite pulses \cite{Levitt86,Torosov11,Torosov18,Torosov19}, while the corresponding control $z$-field varies continuously and monotonically in time. These characteristics differentiate the present study from previous works, where the longitudinal field is also the sole control but it changes discontinuously and non-monotonically \cite{Hegerfeldt13,Hegerfeldt14}.
The present work is expected to find application in the wide spectrum of research fields where AP for two-level systems is exploited.

The structure of the paper is as follows. In the next section we apply the classical RC protocol to the two-level system with detuning control. In Sec. \ref{sec:modified_RC} we present the modification of the RC protocol and formulate the corresponding optimal control problem. In Sec. \ref{sec:optimal_solution} we derive the optimality condition and use it in Sec. \ref{sec:pulse_sequence} to determine the timings of the optimal pulse-sequences. Sec. \ref{sec:conclusion} concludes this work.

\section{Roland-Cerf protocol for a two-level system with detuning control}

\label{sec:RC}

We consider a two-level system with Hamiltonian
\begin{equation}
H(t)=\frac{\Delta(t)}{2}\sigma_z+\frac{\Omega}{2}\sigma_x=\frac{1}{2}
\left[
\begin{array}{cc}
\Delta(t) & \Omega\\
\Omega & -\Delta(t)
\end{array}
\right], \label{Hamiltonian}
\end{equation}
where $\sigma_x, \sigma_z$ are the Pauli spin matrices. The Rabi frequency $\Omega$ ($x$-field) is constant while the time-dependent detuning $\Delta(t)$ ($z$-field) is the control parameter. The instantaneous angle $\theta$ of the total field with respect to $z$-axis is
\begin{equation}
\label{theta}
\cot{\theta(t)}=\frac{\Delta(t)}{\Omega},
\end{equation}
and can also serve as a control parameter instead of the detuning. In terms of $\theta$, Hamiltonian (\ref{Hamiltonian}) is expressed as
\begin{equation}
\label{parametrized_H}
H=\frac{\Omega}{2\sin{\theta}}
\left(\begin{array}{cc}
    \cos\theta & \sin\theta \\ \sin\theta & -\cos\theta
  \end{array}\right).
\end{equation}
If $|\psi\rangle=a_1|0\rangle+a_2|1\rangle$ denotes the state of the system, the probability amplitudes $\bi{a}=(a_1 \; a_2)^T$ obey the equation ($\hbar=1$)
\begin{equation}
\label{Schrodinger_a}
i\dot{\bi{a}}=H\bi{a}.
\end{equation}
The normalized eigenvectors of Hamiltonian (\ref{parametrized_H}) are
\begin{numparts}
\label{eigen}
\begin{eqnarray}
|\phi_{+}(t)\rangle&=&
\left(\begin{array}{c}
    \cos{\frac{\theta}{2}}\\
    \sin{\frac{\theta}{2}}
\end{array}\right),\label{plus}\\
|\phi_{-}(t)\rangle&=&
\left(\begin{array}{c}
    \sin\frac{\theta}{2}\\
    -\cos{\frac{\theta}{2}}
\end{array}\right),\label{minus}
\end{eqnarray}
\end{numparts}
with corresponding eigenvalues
\begin{equation}
\label{eigenenergies}
E_{\pm}(t)=\pm\frac{1}{2}\sqrt{\Delta^2+\Omega^2}=\pm\frac{\Omega}{2\sin{\theta}}.
\end{equation}

\begin{figure}
\centering
\includegraphics[width=.7\linewidth]{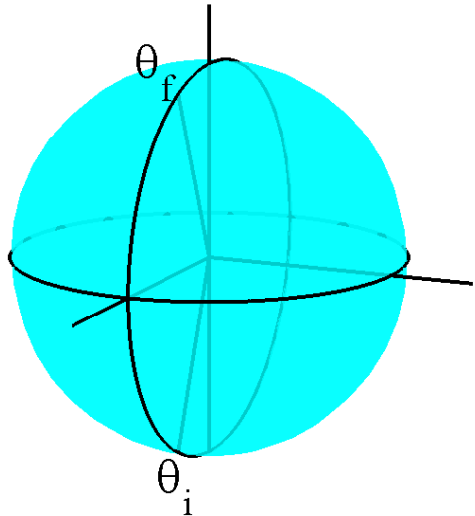}
\caption{Initial and final target states on the Bloch sphere and in the original reference frame, characterized by polar angles $\theta_i, \theta_f$, respectively. The initial and final total fields are also aligned respectively.}
\label{fig:bloch}
\end{figure}

We consider that the detuning starts from a negative value, so the initial angle $\theta_i>\pi/2$ as displayed in Fig. \ref{fig:bloch}, while the system starts from one of the eigenstates (\ref{plus}), (\ref{minus}), with $\theta=\theta_i$. For large initial negative detuning it is $\theta_i\approx \pi$ and $|\phi_{+}\rangle \approx (0 \; 1)^T, |\phi_{-}\rangle \approx (1 \; 0)^T$.
In the traditional AP \cite{Landau32,Zener32}, the detuning is increased linearly with time, until the angle obtains the final value $\theta_f<\pi/2$, see Fig. \ref{fig:bloch}. If the change is slow enough, i.e. for a sufficiently long duration, the system remains in the same eigenstate of the instantaneous Hamiltonian. For large final positive detuning it is $\theta_f\approx 0$ and $|\phi_{+}\rangle \approx (1 \; 0)^T, |\phi_{-}\rangle \approx (0 \; -1)^T$. As at initial and final times each adiabatic state becomes uniquely identified with one of the original states of the system, AP achieves complete population transfer from state $|0\rangle$ to $|1\rangle$ and vice versa. The advantage of the method is its robustness to moderate variation of the system parameters, while its drawback is the long necessary time, which may render it impractical in the presence of decoherence and dissipation. In this article we derive controls, $\Delta(t)$ and $\theta(t)$, which drive the system to the same final eigenstate without following the intermediate adiabatic path.

As a warm up example we present the Roland-Cerf protocol for the two-level system under consideration \cite{Bason12,Malossi13}. In this protocol, the matrix element of the rate of change $dH/dt$ between the eigenstates $|\phi_{\pm}(t)\rangle$,
\begin{equation}
\label{dH_dt}
\langle\phi_{+}(t)|\frac{dH}{dt}|\phi_{-}(t)\rangle=-\frac{\Omega}{2\sin{\theta}}\dot{\theta},
\end{equation}
is taken to be proportional to the square of the instantaneous energy gap,
\begin{equation}
\label{gap}
g(t)=E_{+}(t)-E_{-}(t)=\sqrt{\Delta^2+\Omega^2}=\frac{\Omega}{\sin{\theta}},
\end{equation}
i.e.
\begin{equation}
\label{RC}
\langle\phi_{+}(t)|\frac{dH}{dt}|\phi_{-}(t)\rangle=\frac{1}{2}ug^2(t),
\end{equation}
where $u$ is a \emph{constant} parameter. For $u\ll 1$ the local adiabaticity condition $\langle+|\dot{H}|-\rangle/g^2\ll 1$ is satisfied. For the two-level system, Eq. (\ref{RC}) becomes
\begin{equation}
\label{RC_TL}
\dot{\theta}=-\frac{\Omega}{\sin{\theta}}u,
\end{equation}
which can be easily integrated to give
\begin{equation}
\label{angle_original}
\theta(t)=\cos^{-1}(\cos{\theta_i}+u\Omega t).
\end{equation}
The corresponding detuning can then be obtained from Eq. (\ref{theta}).

The performance of the RC protocol was evaluated numerically in Refs. \cite{Bason12,Malossi13}. In order to evaluate the performance analytically and for arbitrarily large $u$, it is more convenient to work in the adiabatic frame.
By expressing the state of the system in both the original and the adiabatic frames
\begin{equation}
\label{adiabatic_basis}
|\psi\rangle=a_1|0\rangle+a_2|1\rangle=b_1|\phi_{+}\rangle+b_2|\phi_{-}\rangle,
\end{equation}
we obtain the following transformation between the probability amplitudes of the two pictures
\begin{equation}
\label{transformation}
\bi{b}=
\left(\begin{array}{c}
    b_1\\
    b_2
\end{array}\right)
=
\left(\begin{array}{cc}
    \cos{\frac{\theta}{2}} & \sin{\frac{\theta}{2}} \\ \sin{\frac{\theta}{2}} & -\cos{\frac{\theta}{2}}
  \end{array}\right)
\left(\begin{array}{c}
    a_1\\
    a_2
\end{array}\right).
\end{equation}
From Eqs. (\ref{Schrodinger_a}), (\ref{transformation}) we find the following equation for the probability amplitudes in the adiabatic frame
\begin{equation}
\label{Schrodinger_b}
i\dot{\bi{b}}=H_{ad}\bi{b},
\end{equation}
where the Hamiltonian now is
\begin{equation}\label{adiabatic_H}
H_{ad}=\frac{1}{2}
\left(\begin{array}{cc}
    \frac{\Omega}{\sin{\theta}} & -i\dot{\theta} \\ i\dot{\theta} & -\frac{\Omega}{\sin{\theta}}
  \end{array}\right).
\end{equation}

The above equations are simplified if, inspired from Eq. (\ref{RC_TL}), we use a dimensionless rescaled time $\tau$ defined as
\begin{equation}
\label{rescaled_time}
d\tau=\frac{\Omega}{\sin{\theta}}dt.
\end{equation}
For $0<\theta<\pi$, that we consider here, it is $\sin{\theta}>0$ and the rescaling (\ref{rescaled_time}) is well defined.
The equation for $\bi{b}$ becomes
\begin{equation}
\label{rescaled_Schrodinger_b}
i\bi{b}'=H'_{ad}\bi{b},
\end{equation}
where
\begin{equation}
\label{rescaled_adiabatic_H}
H'_{ad}=\frac{1}{2}\sigma_z+\frac{\theta'}{2}\sigma_y=\frac{1}{2}\sigma_z-\frac{u}{2}\sigma_y \, ,
\end{equation}
and $\bi{b}'=d\bi{b}/d\tau$, $\theta'=d\theta/d\tau=-u$ are the derivatives with respect to the rescaled time.
Since Hamiltonian $H'_{ad}$ is constant, from Eqs. (\ref{rescaled_Schrodinger_b}), (\ref{rescaled_adiabatic_H}) we obtain at the final (rescaled) time $\tau=T$ that $\bi{b}(T)=U\bi{b}(0)$, where the unitary transformation $U$ is given by
\begin{equation}
\label{U_simple}
U=e^{-iH'_{ad}T}=e^{-i\frac{1}{2}\omega T(n_z\sigma_z-n_y\sigma_y)}= I\cos{\frac{\omega T}{2}}-i\sin{\frac{\omega T}{2}}(n_z\sigma_z-n_y\sigma_y),
\end{equation}
and
\begin{equation}
\label{parameters}
\omega=\sqrt{1+u^2},\quad n_y=\frac{u}{\omega}=\frac{u}{\sqrt{1+u^2}}, \quad n_z=\frac{1}{\omega}=\frac{1}{\sqrt{1+u^2}}.
\end{equation}

If the system starts in the $|\phi_{+}\rangle$ state, then $\bi{b}(0)=(1 \; 0)^T$. For a perfect AP the system should end up in the same state at the final time $\tau=T$, thus it is sufficient that $b_2(T)=0$. From Eq. (\ref{U_simple}) we obtain the condition $\sin{(\omega T/2)}=0$, such that $U=\pm I$, which leads to
\begin{equation}
\label{shortcut_evolution}
T\sqrt{1+u^2}=2k\pi,\quad k=1,2,\ldots
\end{equation}
During time $T$ the angle should change from $\theta_i$ to $\theta_f$, thus
\begin{equation}
\label{angle_evolution}
\theta_i-\theta_f=-\int_0^{T}\theta'd\tau=uT.
\end{equation}
Combining Eqs. (\ref{shortcut_evolution}) and (\ref{angle_evolution}) we find the solution pairs
\begin{numparts}
\begin{eqnarray}
u_k&=&\frac{\frac{\theta_i-\theta_f}{2k\pi}}{\sqrt{1-\left(\frac{\theta_i-\theta_f}{2k\pi}\right)^2}},\label{uk}\\
T_k&=&2k\pi\sqrt{1-\left(\frac{\theta_i-\theta_f}{2k\pi}\right)^2},\label{Tk}
\end{eqnarray}
\end{numparts}
for $k=1,2,\ldots$ The corresponding durations in the original time $t$ can be found from Eq. (\ref{angle_original}) and they are
\begin{equation}
\label{durations_original}
\tilde{T}_k=\frac{\cos{\theta_f}-\cos{\theta_i}}{u_k}\cdot\frac{1}{\Omega}.
\end{equation}

At this point it is worth mentioning that shortcuts to adiabaticity working for specific durations, like above, have been obtained for quantum teleportation \cite{Oh13} with two control fields playing the role of Stokes and pump pulses in the familiar STIRAP terminology \cite{Kral07,Vitanov17}, as well as for the quantum parametric oscillator \cite{Rezek09,Kosloff17}.

\section{Modified Roland-Cerf protocol as an optimal control problem in the adiabatic reference frame}

\label{sec:modified_RC}

In the previous section we showed that the classical RC protocol, with constant control $u=-d\theta/d\tau$ in the rescaled time, achieves perfect AP for specific durations $T_k$ and amplitudes $u_k$. In the present section we explain how we can generalize this procedure and obtain perfect fidelity for arbitrary durations larger than the lower bound $T_0=\pi$ in the rescaled time, which we derive below. The main idea is to apply a modified RC protocol with time-dependent bounded control $0\leq u(\tau)\leq v$, and then use optimal control theory to obtain the minimum-time pulse-sequence which satisfies all the desired conditions, for specific maximum amplitude $v$. Note that the nonnegativity of $u(\tau)$ assures that the magnetic field angle $\theta$ decreases monotonically from $\theta_i$ to $\theta_f$. On the other hand, as the upper bound $v$ increases, the duration of the optimal pulse-sequence decreases, approaching the limit $T_0=\pi$.

In order to formulate the corresponding optimal control problem in the adiabatic frame, we will use the Bloch equations corresponding to the two-level system (\ref{rescaled_Schrodinger_b}). If we define the new state variables
\begin{numparts}
\label{s}
\begin{eqnarray}
s_x&=&b_1^*b_2+b_1b_2^*,\\
s_y&=&\frac{b_1^*b_2-b_1b_2^*}{i},\\
s_z&=&|b_1|^2-|b_2|^2,
\end{eqnarray}
\end{numparts}
it is not hard to verify that they satisfy the following equations
\begin{numparts}
\label{system}
\begin{eqnarray}
\dot{s}_x&=&-s_y-us_z,\\
\dot{s}_y&=&s_x,\\
\dot{s}_z&=&us_x,
\end{eqnarray}
\end{numparts}
or, in a more compact form
\begin{equation}
\label{compact_system}
\dot{\bi{s}}=(Z-uY)\bi{s},
\end{equation}
where $\bi{s}=(s_x, s_y, s_z)^T$ and
\begin{equation}
\label{XYZ}
X=
\left (
\begin{array}{ccc}
0 & 0 & 0\\
0 & 0 & -1\\
0 & 1 & 0
\end{array}
\right ),
\quad
Y=
\left (
\begin{array}{ccc}
0 & 0 & 1\\
0 & 0 & 0\\
-1 & 0 & 0
\end{array}
\right ),
\quad
Z=
\left (
\begin{array}{ccc}
0 & -1 & 0\\
1 & 0 & 0\\
0 & 0 & 0
\end{array}
\right ).
\quad
\end{equation}
Since the matrices in Eq. (\ref{XYZ}) are antisymmetric, the system equation (\ref{compact_system}) can take the form
\begin{equation}
\label{system_cross_product}
\dot{\bi{s}}=(\hat{\bi{z}}-u\hat{\bi{y}})\times\bi{s},
\end{equation}
where $\times$ denotes the vector cross product and
\begin{equation}
\label{xyz}
\hat{\bi{x}}=
\left (
\begin{array}{c}
1 \\
0 \\
0
\end{array}
\right ),
\quad
\hat{\bi{y}}=
\left (
\begin{array}{c}
0 \\
1 \\
0
\end{array}
\right ),
\quad
\hat{\bi{z}}=
\left (
\begin{array}{c}
0 \\
0 \\
1
\end{array}
\right )
\quad
\end{equation}
are the axes unit vectors.

We can now formulate the optimal control problem for system (\ref{compact_system}) or (\ref{system_cross_product}). Starting from the north pole $\bi{s}=(0, 0, 1)^T$, we would like to find the bounded control $0\leq u(\tau)\leq v$ with specified area $\int_0^Tu(\tau)d\tau=\theta_i-\theta_f$ which minimizes the time $T=\int_0^T1d\tau$ needed to return to the starting point. In the following section we analyze the solutions to this problem using optimal control theory. Before doing so, we explain how is obtained the lower bound $T_0=\pi$ (in the rescaled time) of the pulse-sequence duration.

In the \emph{original reference frame} (not the adiabatic), we consider an instantaneous change in the total field from $\theta=\theta_i$ to $\theta=\bar{\theta}=(\theta_i+\theta_f)/2$, i.e. in the middle of the arc connecting the initial and target states. The corresponding detuning is $\Delta=\Omega\cot{\bar{\theta}}$ and the total field is $\sqrt{\Delta^2+\Omega^2}=\Omega/\sin{\bar{\theta}}$. Under the influence of this constant field for duration
\begin{equation}
\label{original_bound}
\tilde{T}_0=\sin{\bar{\theta}}\frac{\pi}{\Omega}=\sin{\frac{\theta_i+\theta_f}{2}}\frac{\pi}{\Omega},
\end{equation}
the Bloch vector is rotated from $(\phi=0,\theta_i)$ to $(\phi=0,\theta_f)$. After the completion of this half circle, the total field is changed again instantaneously from $\theta=\bar{\theta}$ to $\theta=\theta_f$. Since $\theta=\bar{\theta}$ during this evolution, except the (measure zero) initial and final instants, Eq. (\ref{rescaled_time}) becomes $d\tau=\Omega dt/\sin{\bar{\theta}}$ and the corresponding duration in the rescaled time is thus
\begin{equation}
\label{rescaled_bound}
T_0=\frac{\Omega \tilde{T}_0}{\sin{\bar{\theta}}}=\pi.
\end{equation}
We finally point out that the corresponding quantum speed limit (in the original time) is $\tilde{T}_{qsl}=(\theta_i-\theta_f)/\Omega$, as obtained in Ref. \cite{Calarco09} and formally proved in Ref. \cite{Hegerfeldt13}, see also Refs. \cite{Bason12,Malossi13,Poggi13}, but is derived using infinite values of the detuning which implement instantaneous rotations around $z$-axis, while angle $\theta$ changes non-monotonically. More realistic speed limits have been obtained for bounded detuning \cite{Hegerfeldt14}, but their implementation also requires discontinuous and non-monotonic changes of the magnetic field angle. On the contrary, the bounds in Eqs. (\ref{original_bound}), (\ref{rescaled_bound}) are obtained with finite detuning values and a monotonic change of $\theta$ (decrease for $\theta_i>\theta_f$). For $\theta_i\approx\pi$ and $\theta_f\approx 0$, it is $\tilde{T}_{qsl}\approx \tilde{T}_0\approx \pi/\Omega$, as derived in \cite{Boscain02}.

\section{Analysis of the optimal solution}

\label{sec:optimal_solution}

Let $\blambda=(\lambda_x, \lambda_y, \lambda_z)$ be the time-dependent \emph{row} vector of Lagrange multipliers corresponding to system equations and $\mu$ the \emph{constant} multiplier corresponding to the integral condition for the pulse area. The control Hamiltonian for the previously formulated problem incorporates aside the cost (time) both the integral condition and the system equation
\begin{eqnarray}
\label{Hc}
H_c&=&1+\mu u+\blambda\cdot\dot{\bi{s}}\nonumber\\
   &=&1+(\mu-\blambda\cdot Y\bi{s})u+\blambda\cdot Z\bi{s}\nonumber\\
   &=&1+(\mu+\lambda_zs_x-\lambda_xs_z)u+\lambda_ys_x-\lambda_xs_y.
\end{eqnarray}
Using Hamilton's equations $\dot{\lambda}_{\alpha}=-\partial H_c/\partial s_{\alpha}$, $\alpha=x,y,z$, we find the following equation for the adjoint variables
\begin{equation}
\label{adjoint_equation}
\dot{\blambda}=-\blambda(Z-uY).
\end{equation}
Note that multiplier $\mu$ is constant since the corresponding coordinate, angle $\theta$, is cyclic.

\begin{figure}[t]
\centering
\includegraphics[width=0.6\linewidth]{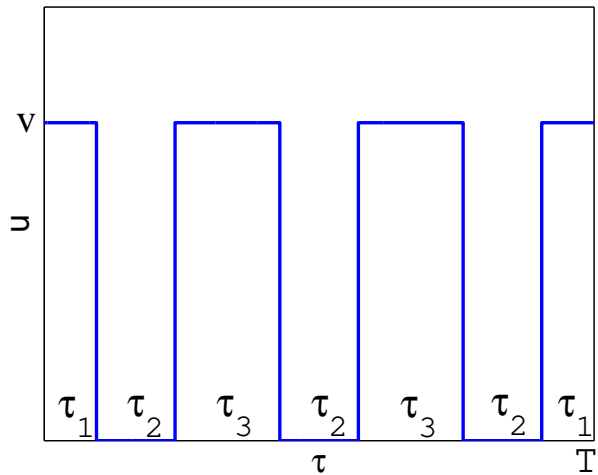}
\caption{Candidate optimal pulse-sequences $u(\tau)$ in the rescaled time $\tau$. The initial and final ``on" pulses have the same duration $\tau_1$, all the intermediate ``off" pulses have the same duration $\tau_2$, while all the intermediate ``on" pulses have the same duration $\tau_3$. The middle pulse can be ``off", as in this figure, or ``on". The total duration of the sequence is $T$.}
\label{fig:candidate_sequence}
\end{figure}

According to Pontryagin Maximum Principle \cite{Pontryagin}, the optimal control $0\leq u(\tau)\leq v$ is chosen to minimize $H_c$. If we define the functions
\begin{numparts}
\begin{eqnarray}
\phi_x&=&\lambda_ys_z-\lambda_zs_y,\label{phix}\\
\phi_y&=&\mu+\lambda_zs_x-\lambda_xs_z,\label{phiy}\\
\phi_z&=&\lambda_ys_x-\lambda_xs_y,\label{phiz}
\end{eqnarray}
\end{numparts}
then the control Hamiltonian can be expressed as $H_c=1+\phi_yu+\phi_z$. Obviously, $H_c$ is a linear function of the bounded control $0\leq u\leq v$ with coefficient $\phi_y$, the so-called switching function. The optimal $u$ minimizing $H_c$ is $u=0$ for $\phi_y>0$ and $u=v$ for $\phi_y<0$. If $\phi_y=0$ for some finite time interval, then $u$ takes some intermediate value which cannot be found from Maximum Principle. However, if $\phi_y(\tau)=0$ and $\dot{\phi}_y(\tau)\neq0$, then at time $\tau$ the control switches between its boundary values and we call this a bang-bang switch. In the present article we concentrate on bang-bang solutions, i.e. pulse-sequences of the form ``on-off-on-...-on-off-on", where $u(\tau)$ alternates between $0$ and its maximum value $v$, as displayed in Fig. \ref{fig:candidate_sequence}. For each value of parameter $v$ we will find the timings of the corresponding optimal pulse-sequence.

\begin{figure}
\centering
\includegraphics[width=.7\linewidth]{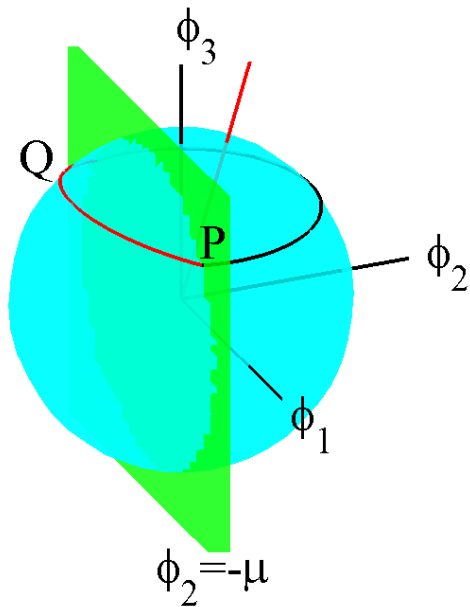}
\caption{Trajectory of vector $\bphi$ for $u=0$, black horizontal arc corresponding to a rotation around the vertical black axis, and $u=v$, inclined red arc corresponding to a rotation around the tilted red axis. On the switching plane $\phi_2=-\mu$ the control changes from the one boundary value to the other and the evolution repeats itself.}
\label{fig:proof}
\end{figure}

We start by showing geometrically that in the optimal bang-bang pulse-sequence all the ``off" pulses have the same duration, say $\tau_2$, and all the intermediate ``on" pulses (i.e. aside the first and the last) have the same duration, say $\tau_3$. Using the equations for the state and adjoint variables we can show that the vector $\bphi=(\phi_1, \phi_2, \phi_3)^T=(\phi_x, \phi_y-\mu, \phi_z)^T$ obeys the following equation
\begin{equation}
\label{phi}
\dot{\bphi}=(Z+uY)\bphi,
\end{equation}
or
\begin{equation}
\label{phi_vector}
\dot{\bphi}=(\hat{\bi{z}}+u\hat{\bi{y}})\times\bphi,
\end{equation}
if we use vectors instead of antisymmetric matrices. From the last equation it is obvious that the motion of $\bphi$ is restricted on a sphere,
\begin{equation}
\label{sphere}
\phi_1^2+\phi_2^2+\phi_3^2=\phi_x^2+(\phi_y-\mu)^2+\phi_z^2=\mbox{constant}.
\end{equation}
Now suppose that at time $\tau$ there is a switching from $u=v$ to $u=0$. This means that $\phi_y(\tau)=0$, which also implies $\phi_2(\tau)=-\mu$, thus the switching point $P(\bar{\phi}_1,-\mu,\bar{\phi}_3)$ lies on the plane $\phi_2=-\mu$, shown with green color in Fig. \ref{fig:proof}, while $\bar{\phi}_1,\bar{\phi}_3$ denote the other two coordinates of $P$. The control $u=0$ is applied for duration $\tau_2$ and $\bphi$ is rotated around $z$-axis along the horizontal black arc displayed in Fig. \ref{fig:proof}. Note that during this interval it is $\phi_2>-\mu\Rightarrow\phi_y>0$, thus $u=0$ minimizes indeed the control Hamiltonian. At time $\tau+\tau_2$ the trajectory intersects the switching plane $\phi_2=-\mu$ at point $Q(-\bar{\phi}_1,-\mu,\bar{\phi}_3)$, the symmetric of $P$ with respect to the $\phi_2\phi_3$-plane. Since we consider bang-bang pulse-sequences, the control switches from $u=0$ to $u=v$. Vector $\bphi$ is now rotated around the (red) axis $\bi{n}=\hat{\bi{z}}+u\hat{\bi{y}}$ for duration $\tau_3$, along the inclined red arc shown in Fig. \ref{fig:proof}. During this time interval it is $\phi_2<-\mu\Rightarrow\phi_y<0$, thus $u=v$ minimizes indeed the control Hamiltonian. At time $\tau+\tau_2+\tau_3$ the trajectory meets again the switching plane $\phi_2=-\mu$; we will show that this intersection takes place at point $P$. During the rotation around axis $\bi{n}=\hat{\bi{z}}+u\hat{\bi{y}}$, the inner product $\bphi\cdot\bi{n}=\phi_3+v\phi_2$ is constant. But $\phi_2(\tau+\tau_2+\tau_3)=-\mu=\phi_2(\tau+\tau_2)$, thus $\phi_3(\tau+\tau_2+\tau_3)=\phi_3(\tau+\tau_2)=\bar{\phi}_3$. Since the motion is restricted on the sphere (\ref{sphere}), we easily deduce that $\phi_1(\tau+\tau_2+\tau_3)=\bar{\phi}_1$. The trajectory thus intersects the switching plane at the point $P(\bar{\phi}_1,-\mu,\bar{\phi}_3)$, and the evolution is repeated for all the subsequent ``off" and intermediate ``on" pulses. The conclusion is that all the ``off" pulses have the same duration $\tau_2$, and all the intermediate ``on" pulses have the same duration $\tau_3$.

The initial and final ``on" pulses can have different durations than $\tau_3$, corresponding to incomplete traversals of the red arc shown in Fig. \ref{fig:proof}. Since the system (\ref{compact_system}) starts from and returns to the same point, the north pole, for symmetry reasons we take the initial and final ``on" pulses to have the same duration $\tau_1$. Thus, we consider candidate optimal pulse-sequences of the form shown in Fig. \ref{fig:candidate_sequence} and the optimization takes place within this subset. In the following we use geometric optimal control \cite{Schattler12} to derive a relation between the pulse durations $\tau_1, \tau_2, \tau_3$. This relation will be exploited in the next section, along with the integral condition for the pulse area and the condition that the system should return to the north pole at the final time, in order to obtain these durations when the maximum control amplitude $v$ is given. In the rest of this section we particularly use the theory developed in Ref. \cite{Sussmann87}, as specified for the two-level quantum system in Refs. \cite{Boscain05,Boscain06}, while we adopt it to incorporate the pulse area condition.

Observe from the second line of Eq. (\ref{Hc}) that the switching function can be expressed as $\phi_y=\mu-\blambda\cdot Y\bi{s}$, thus at a switching time $\tau$ it holds
\begin{equation}
\phi_y(\tau)=0\Rightarrow\blambda(\tau)\cdot Y\bi{s}(\tau)=\mu.
\end{equation}
For the first three switchings at times $\tau=\tau_1, \tau_1+\tau_2, \tau_1+\tau_2+\tau_3$ we have
\begin{equation}
\label{consecutive_switchings}
\fl\blambda(\tau_1)\cdot Y\bi{s}(\tau_1)=\blambda(\tau_1+\tau_2)\cdot Y\bi{s}(\tau_1+\tau_2)=\blambda(\tau_1+\tau_2+\tau_3)\cdot Y\bi{s}(\tau_1+\tau_2+\tau_3)=\mu.
\end{equation}
We will express the first and third terms of the above equation at the middle time $\tau=\tau_1+\tau_2$. During the interval $\tau_1<\tau\leq\tau_1+\tau_2$ the control is $u=0$.
From Eqs. (\ref{compact_system}), (\ref{adjoint_equation}) for the state and adjoint variables $\bi{s}, \blambda$ we have $\bi{s}(\tau_1)=e^{-\tau_2Z}\bi{s}(\tau_1+\tau_2), \blambda(\tau_1)=\blambda(\tau_1+\tau_2)e^{\tau_2Z}$, thus
\begin{equation}
\label{l_tau1}
\blambda(\tau_1)\cdot Y\bi{s}(\tau_1)=\blambda(\tau_1+\tau_2)\cdot Y_1\bi{s}(\tau_1+\tau_2),
\end{equation}
where
\begin{equation}
\label{Y1}
Y_1=e^{\tau_2Z}Ye^{-\tau_2Z}=\cos{\tau_2}Y-\sin{\tau_2}X.
\end{equation}
Note that in the derivation of the last equation we have used the commutation relations $[X,Y]=Z, [Y,Z]=X, [Z,X]=Y$.
Analogously, during the interval $\tau_1+\tau_2<\tau\leq\tau_1+\tau_2+\tau_3$ the control is $u=v$, thus $\bi{s}(\tau_1+\tau_2+\tau_3)=e^{\tau_3(Z-vY)}\bi{s}(\tau_1+\tau_2), \blambda(\tau_1+\tau_2+\tau_3)=\blambda(\tau_1+\tau_2)e^{-\tau_3(Z-vY)}$ and
\begin{equation}
\label{l_tau3}
\blambda(\tau_1+\tau_2+\tau_3)\cdot Y\bi{s}(\tau_1+\tau_2+\tau_3)=\blambda(\tau_1+\tau_2)\cdot Y_2\bi{s}(\tau_1+\tau_2),
\end{equation}
where
\begin{eqnarray}
\label{Y2}
Y_2&=&e^{-\tau_3(Z-vY)}Ye^{\tau_3(Z-vY)}\\
&=&\frac{\sin{\omega\tau_3}}{\omega}X+\frac{u^2+\cos{\omega\tau_3}}{\omega^2}Y-\frac{u(1-\cos{\omega\tau_3})}{\omega^2}Z
\end{eqnarray}
and
\begin{equation}
\label{w}
\omega=\sqrt{1+v^2}.
\end{equation}
Using Eqs. (\ref{l_tau1}), (\ref{l_tau3}), Eq. (\ref{consecutive_switchings}) becomes
\begin{equation}
\label{same_switchings}
\fl\blambda(\tau_1+\tau_2)\cdot Y_1\bi{s}(\tau_1+\tau_2)=\blambda(\tau_1+\tau_2)\cdot Y\bi{s}(\tau_1+\tau_2)=\blambda(\tau_1+\tau_2)\cdot Y_2\bi{s}(\tau_1+\tau_2)=\mu.
\end{equation}
Since $Y, Y_1, Y_2$ are antisymmetric matrices, the above equation can be expressed using the corresponding vectors $\hat{\bi{y}}$,
\begin{eqnarray}
\bi{y_1}&=&\cos{\tau_2}\hat{\bi{y}}-\sin{\tau_2}\hat{\bi{x}},\\
\bi{y_2}&=&\frac{\sin{\omega\tau_3}}{\omega}\hat{\bi{x}}+\frac{u^2+\cos{\omega\tau_3}}{\omega^2}\hat{\bi{y}}-\frac{u(1-\cos{\omega\tau_3})}{\omega^2}\hat{\bi{z}},
\end{eqnarray}
as
\begin{equation}
\label{same_switchings_vector}
\fl\blambda(\tau_1+\tau_2)\cdot \bi{y_1}\times\bi{s}(\tau_1+\tau_2)=\blambda(\tau_1+\tau_2)\cdot \hat{\bi{y}}\times\bi{s}(\tau_1+\tau_2)=\blambda(\tau_1+\tau_2)\cdot \bi{y_2}\times\bi{s}(\tau_1+\tau_2)=\mu,
\end{equation}
from which we obtain
\begin{equation}
\label{mixed_products}
\fl\blambda(\tau_1+\tau_2)\cdot (\hat{\bi{y}}-\bi{y_1})\times\bi{s}(\tau_1+\tau_2)=\blambda(\tau_1+\tau_2)\cdot (\hat{\bi{y}}-\bi{y_2})\times\bi{s}(\tau_1+\tau_2)=0.
\end{equation}
Note that $\blambda(\tau_1+\tau_2)\neq \mathbf{0}$, since otherwise $\mu=0$ from Eq. (\ref{same_switchings_vector}) and the homogeneous equation (\ref{adjoint_equation}) would imply $\blambda=\mathbf{0}$ for all times, i.e. all the multipliers would be zero, something which contradicts Maximum Principle \cite{Pontryagin}. Now, according to the above equation, the nonzero vector $\blambda(\tau_1+\tau_2)$ is perpendicular to the cross product $(\hat{\bi{y}}-\bi{y_1})\times\bi{s}(\tau_1+\tau_2)$, thus the vectors $\blambda(\tau_1+\tau_2),\hat{\bi{y}}-\bi{y_1},\bi{s}(\tau_1+\tau_2)$ are coplanar. Analogously we show that the vectors $\blambda(\tau_1+\tau_2),\hat{\bi{y}}-\bi{y_2},\bi{s}(\tau_1+\tau_2)$ are also coplanar. The conclusion is that the vectors $\hat{\bi{y}}-\bi{y_1},\hat{\bi{y}}-\bi{y_2},\bi{s}(\tau_1+\tau_2)$ are coplanar, where
\begin{eqnarray}
\label{st1t2}
\bi{s}(\tau_1+\tau_2)&=&e^{\tau_2Z}e^{\tau_1(Z-uY)}\bi{s}(0)\nonumber\\
&=&
\left(
\begin{array}{c}
-n_y\sin{\omega\tau_1}\cos{\tau_2}+n_y^2(1-\cos{\omega\tau_1})\sin{\tau_2}\\
-n_y\sin{\omega\tau_1}\sin{\tau_2}-n_y^2(1-\cos{\omega\tau_1})\cos{\tau_2}\\
n_z^2+n_y^2\cos{\omega\tau_1}
\end{array}
\right)
\end{eqnarray}
and
\begin{equation}
n_y=\frac{v}{\omega}=\frac{v}{\sqrt{1+v^2}},\quad n_z=\frac{1}{\omega}=\frac{1}{\sqrt{1+v^2}}.
\end{equation}
Three coplanar vectors are linearly dependent, thus
\begin{equation*}
\mbox{det}(\hat{\bi{y}}-\bi{y_1}, \hat{\bi{y}}-\bi{y_2}, \bi{s}(\tau_1+\tau_2))=0,
\end{equation*}
leading to
\begin{equation}
\label{optimality}
A\sin{\tau_2}+B(1-\cos{\tau_2})=0,
\end{equation}
where
\begin{numparts}
\label{AB}
\begin{eqnarray}
A&=&(1-\cos{\omega\tau_3})[n_z+n_y^2(n_y-n_z)(1-\cos{\omega\tau_1})],\label{A}\\
B&=&n_y^2\sin{\omega\tau_1}+n_z^2\sin{\omega\tau_3}+n_y^2\sin{[\omega(\tau_3-\tau_1)]}.\label{B}
\end{eqnarray}
\end{numparts}
Eq. (\ref{optimality}) is the optimality condition between the pulse durations $\tau_1, \tau_2, \tau_3$, for a given maximum control amplitude $v$ included in $\omega, n_y, n_z$.

\section{Optimal pulse-sequences}

\label{sec:pulse_sequence}

In this section we use optimality condition (\ref{optimality}), along with the pulse area condition $\int_0^Tu(\tau)d\tau=\theta_i-\theta_f$ and the final condition that the system returns to the north pole, in order to obtain the timings $\tau_1, \tau_2, \tau_3$ for the optimal pulse-sequences. Let us consider a pulse-sequence $u(\tau)$ containing $m$ ``off" pulses, where $m=1,2,\ldots$ is a positive integer. Since the ``on" pulses have constant amplitude $v$, the total change in the angle $\theta$ is
\begin{equation*}
\theta_i-\theta_f=v[2\tau_1+(m-1)\tau_3],
\end{equation*}
thus
\begin{equation}
\label{first_relation}
\tau_1=\frac{1}{2}\left [\frac{\theta_i-\theta_f}{v}-(m-1)\tau_3\right].
\end{equation}
Next, observe that Eq. (\ref{optimality}) can be solved with respect to $\tau_2$
\begin{equation}
\label{second_relation}
\tau_2=2\cot^{-1}\left(-\frac{B}{A}\right),
\end{equation}
where note from Eqs. (\ref{A}), (\ref{B}) that $A, B$ are functions of $\tau_1, \tau_3$ only. Since $\tau_1$ is expressed as a function of $\tau_3$ in Eq. (\ref{first_relation}), obviously $\tau_2$ can also be expressed as a function of $\tau_3$ only.

The last relation that we need is derived from the requirement that the system should return to the north pole. Instead of the Bloch system (\ref{system_cross_product}), it is more convenient to use system (\ref{rescaled_Schrodinger_b}), for which the corresponding final condition is that it should return to the adiabatic state $|\phi_{+}\rangle$ at the final time $\tau=T$. Under the piecewise constant pulse-sequence $u(\tau)$, the propagator $U$ connecting the initial and final states, $\bi{b}(T)=U\bi{b}(0)$, can be expressed as
\begin{equation}
\label{total_propagator}
U=U_1W_2U_3\ldots W_2\,\mbox{or}\,U_3\ldots U_3W_2U_1,
\end{equation}
where $U_j$, $j=1, 3$, is given by
\begin{equation}
\label{U}
U_j=e^{-iH'_{ad}\tau_j}=e^{-i\frac{1}{2}\omega \tau_j(n_z\sigma_z-n_y\sigma_y)}=I\cos{\frac{\omega\tau_j}{2}}-i\sin{\frac{\omega\tau_j}{2}}(n_z\sigma_z-n_y\sigma_y),
\end{equation}
and
\begin{equation}
\label{W}
W_2=e^{-i\frac{1}{2}\tau_2\sigma_z}=I\cos{\frac{\tau_2}{2}}-i\sin{\frac{\tau_2}{2}}\sigma_z.
\end{equation}
The propagator in the middle of (\ref{total_propagator}) is $W_2$ or $U_3$, depending on the corresponding middle pulse.
Using the expressions for $U_1, W_2, U_3$ and the following property of Pauli matrices
\begin{equation}
\label{Pauli}
\sigma_a\sigma_b=\delta_{ab}I+i\epsilon_{abc}\sigma_c,
\end{equation}
where $a,b,c$ can be any of $x,y,z$, $\delta_{ab}$ is the Kronecker delta and $\epsilon_{abc}$ is the Levi-Civita symbol,
we can express the propagator $U$ as a linear combination of $\sigma_a$ and the identity $I$,
\begin{equation}
\label{propagator}
U=a_II+a_x\sigma_x+a_y\sigma_y+a_z\sigma_z.
\end{equation}
The coefficients of the matrices in the above expression are functions of the pulse-sequence parameters.

In the appendix we show that $a_x=0$. Now observe that $I, \sigma_z$ are diagonal. Since $a_x=0$ in Eq. (\ref{propagator}), if we set $a_y=0$ then $U$ is also diagonal. In this case, starting from $\bi{b}(0)=(1 \; 0)^T$ we find for the final state $\bi{b}(T)=U\bi{b}(0)$ that $b_2(T)=0$, and the system returns to the initial adiabatic state. The relation
\begin{equation}
\label{third_relation}
a_{y,m}(\tau_1,\tau_2,\tau_3,v)=0,
\end{equation}
along with Eqs. (\ref{first_relation}), (\ref{second_relation}), will be used for the determination of the pulse-sequence timing parameters $\tau_1,\tau_2,\tau_3$. The subscript $m$ denotes that $a_y$ has a different functional form for pulse-sequences with different number $m$ of ``off" pulses. Following the procedure described in the appendix, we have found $a_{y,m}$ for $m=1, 2, 3$,
\begin{eqnarray}
\label{ay1}
\fl a_{y,1}=\frac{1}{2}\mbox{Tr}(\sigma_yU)=\frac{1}{2}\mbox{Tr}(\sigma_yU_1W_2U_1)=\frac{1}{2}\mbox{Tr}(U_1\sigma_yU_1W_2)\nonumber\\
\fl=2i n_y\sin{(\omega \tau_1/2)}\big[\cos{(\omega \tau_1/2)}\cos{(\tau_2/2)}-n_z\sin{(\omega \tau_1/2)}\sin{(\tau_2/2)}\big],\nonumber\\
\end{eqnarray}
\begin{eqnarray}
\fl
\label{ay2}
a_{y,2}=\frac{1}{2}\mbox{Tr}(\sigma_yU)=\frac{1}{2}\mbox{Tr}(\sigma_yU_1W_2U_3W_2U_1)=\frac{1}{2}\mbox{Tr}(W_2U_1\sigma_yU_1W_2U_3)\nonumber\\
\fl=in_y\cos{(\omega \tau_3/2)}\big[\sin{\omega \tau_1}\cos{\tau_2}-n_z\sin{\tau_2}(1-\cos{\omega \tau_1})\big]\nonumber\\
\fl + in_y\sin{(\omega \tau_3/2)}\bigg\{\cos{\omega \tau_1}+n_z\big[-\sin{\omega \tau_1}\sin{\tau_2}+n_z(1-\cos{\omega \tau_1})(1-\cos{\tau_2})\big]\bigg\},\nonumber\\
\end{eqnarray}
\fl
\begin{eqnarray}
\label{ay3}
\fl a_{y,3}=\frac{1}{2}\mbox{Tr}(\sigma_yU)=\frac{1}{2}\mbox{Tr}(\sigma_yU_1W_2U_3W_2U_3W_2U_1)=\frac{1}{2}\mbox{Tr}(U_3W_2U_1\sigma_yU_1W_2U_3W_2)\nonumber\\
\fl=in_y\big[\cos{(\tau_2/2)}\cos{\omega \tau_3}-n_z\sin{(\tau_2/2)}\sin{\omega \tau_3}\big]\nonumber \\
\fl \times \big[\sin{\omega \tau_1}\cos{\tau_2}-n_z\sin{\tau_2}(1-\cos{\omega \tau_1})\big]\nonumber\\
\fl +in_y\big[n_z\cos{(\tau_2/2)}\sin{\omega \tau_3}+\sin{(\tau_2/2)}(n_y^2+n_z^2\cos{\omega \tau_3})\big]\nonumber \\
\fl \times \big[-\sin{\omega \tau_1}\sin{\tau_2}+n_z(1-\cos{\omega \tau_1})(1-\cos{\tau_2})\big]\nonumber\\
\fl +in_y\big[\cos{\omega \tau_1}\cos{(\tau_2/2)}\sin{\omega \tau_3}-n_z\sin{(\tau_2/2)}(1-\cos{\omega \tau_1}\cos{\omega \tau_3})\big].
\end{eqnarray}

Observe that for $m=1$, i.e. the simplest ``on-off-on" pulse-sequence, there are no intermediate ``on" pulses, thus $\tau_3=0$. In this case, Eqs. (\ref{A}), (\ref{B}) give $A=B=0$, and the optimality condition (\ref{optimality}) is automatically satisfied. From Eq. (\ref{first_relation}) we have $\tau_1=(\theta_i-\theta_f)/(2v)$, while equation $a_{y,1}=0$ becomes a transcendental equation for unknown duration $\tau_2$. For $m>1$, using Eqs. (\ref{first_relation}), (\ref{second_relation}) in Eq. (\ref{third_relation}), we end up with a transcendental equation for $\tau_3$. For each value of the maximum control amplitude $v>0$, the transcendental equations corresponding to different $m$ may or may not have solutions. For each solution we find the total duration $T$ of the corresponding pulse-sequence and compare the results. The pulse-sequence with the minimum $T$ is the optimal one for the specific value of $v$.

\begin{figure*}[t]
 \centering
		\begin{tabular}{cc}
     	\subfigure[$\ $]{
	            \label{fig:duration_rescaled}
	            \includegraphics[width=.45\linewidth]{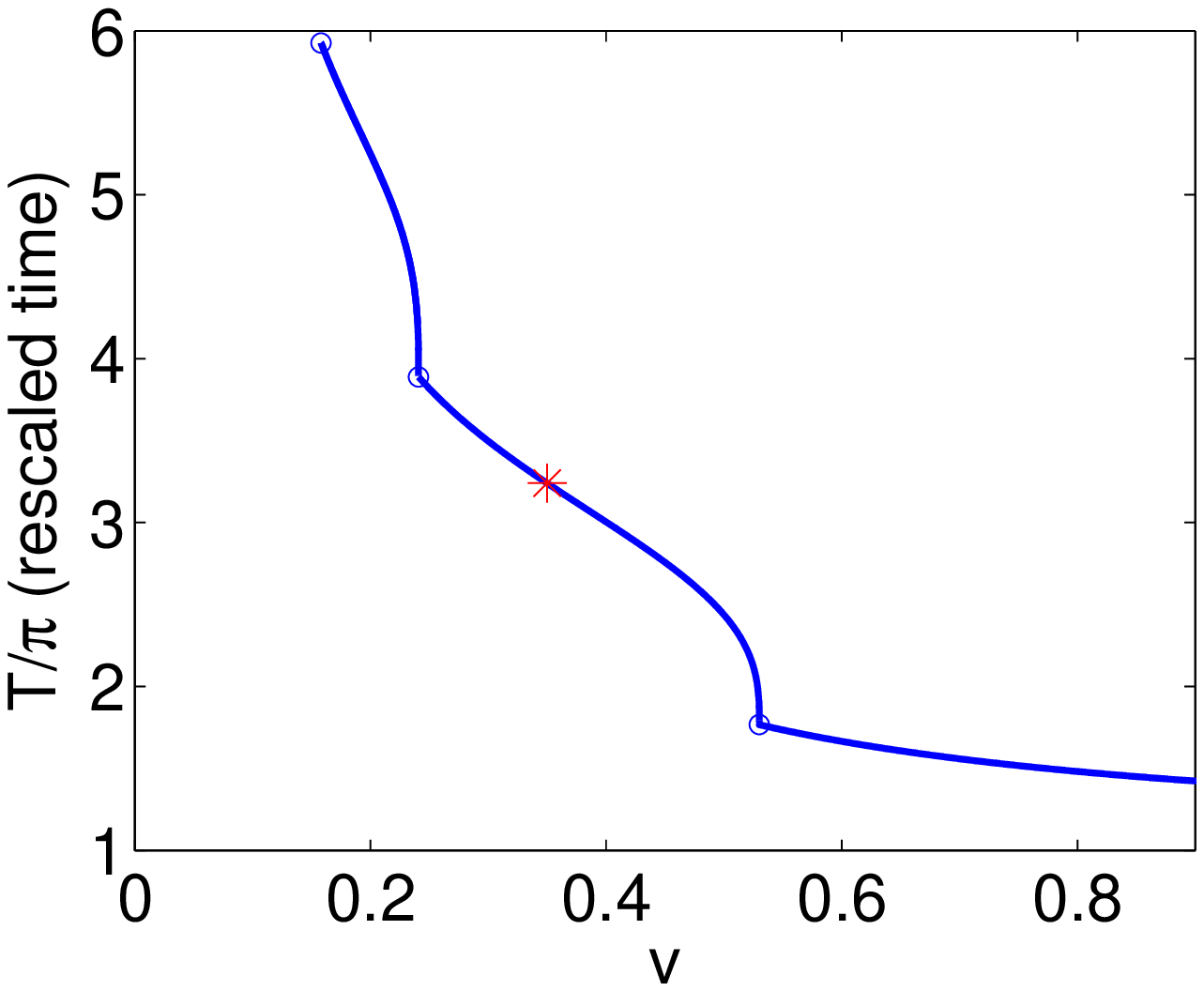}} &
       \subfigure[$\ $]{
	            \label{fig:duration_original}
	            \includegraphics[width=.45\linewidth]{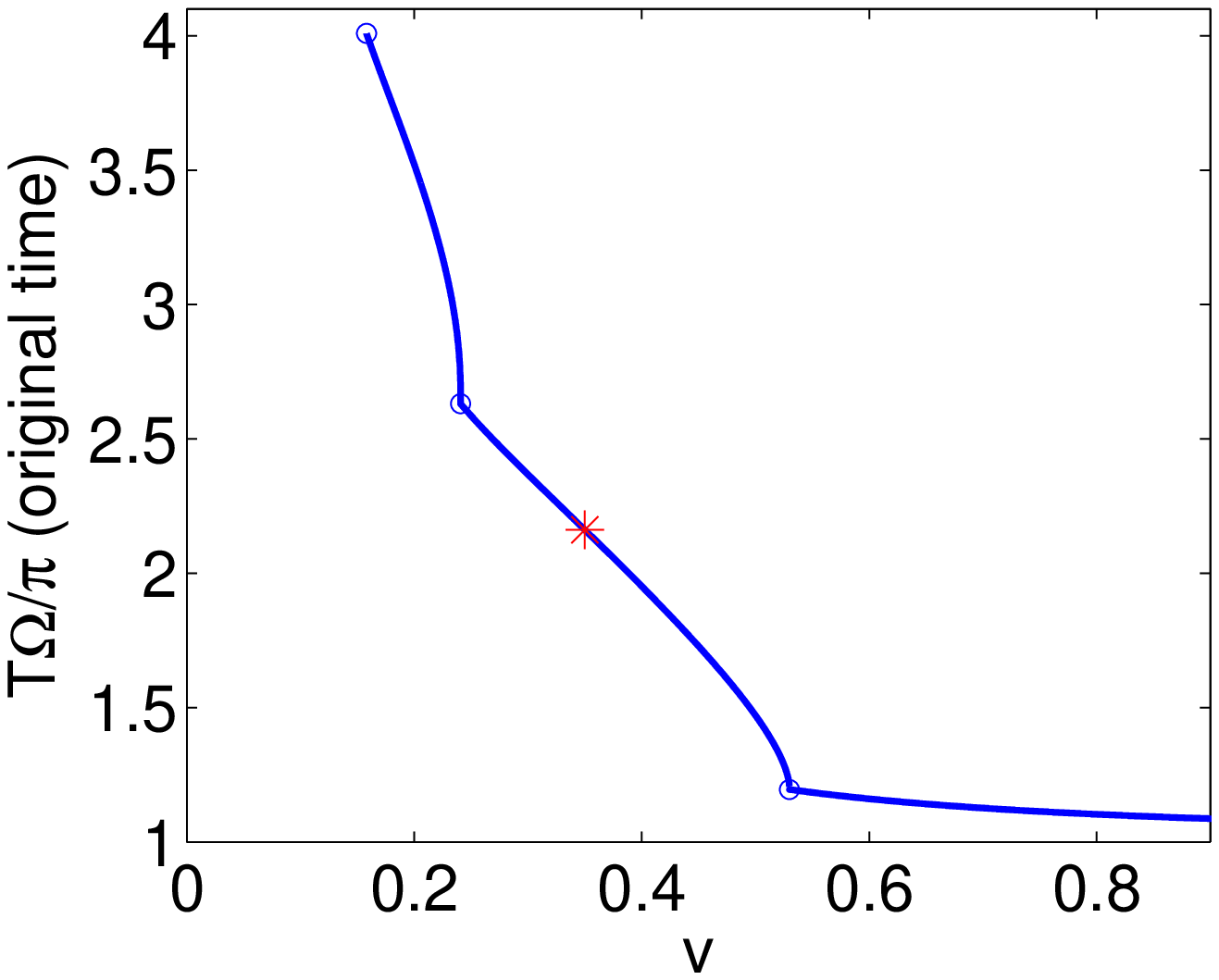}}
		\end{tabular}
\caption{Duration of the optimal pulse-sequence as a function of the maximum control amplitude $v$, both in the rescaled time (a) and the original time (b), for $\theta_f=\tan^{-1}(1/10)$ and $\theta_i=\pi-\theta_f$. The diagrams display a stairway-like form, where the circles separating the steps are the points where the original Roland-Cerf protocol, with constant control $u(\tau)=v$, is optimal. On the first step from the right (larger values of $v$), the optimal pulse-sequence has the simple ``on-off-on" form. Note that for large values of $v$ the optimal duration tends to the limiting value $\pi$. On the second step, the optimal pulse-sequence changes to ``on-off-on-off-on", on the third step becomes ``on-off-on-off-on-off-on", and so forth.}
\label{fig:durations}
\end{figure*}

\begin{figure*}[t]
 \centering
		\begin{tabular}{cc}
     	\subfigure[$\ $]{
	            \label{fig:resonance}
	            \includegraphics[width=.45\linewidth]{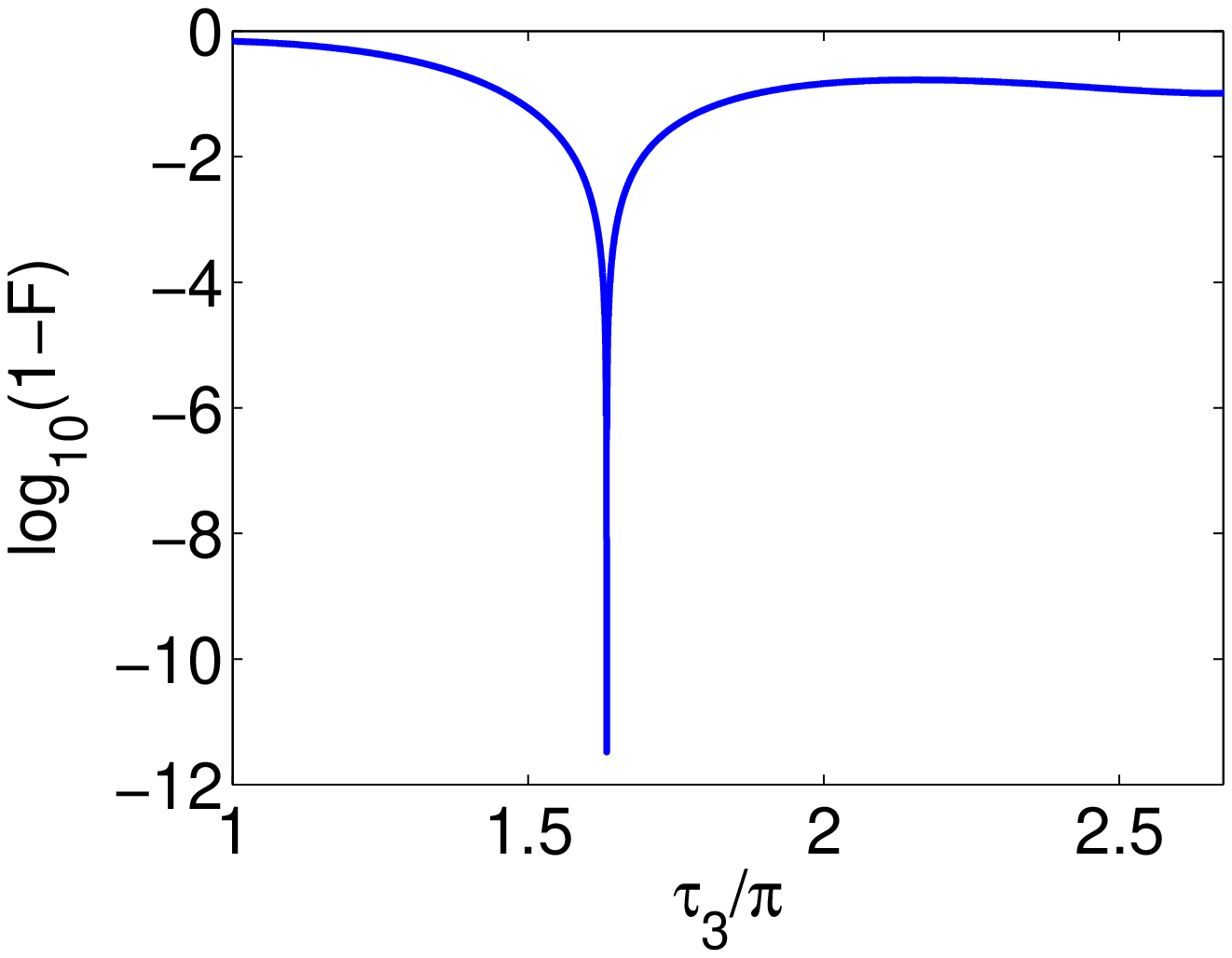}} &
       \subfigure[$\ $]{
	            \label{fig:sequence}
	            \includegraphics[width=.45\linewidth]{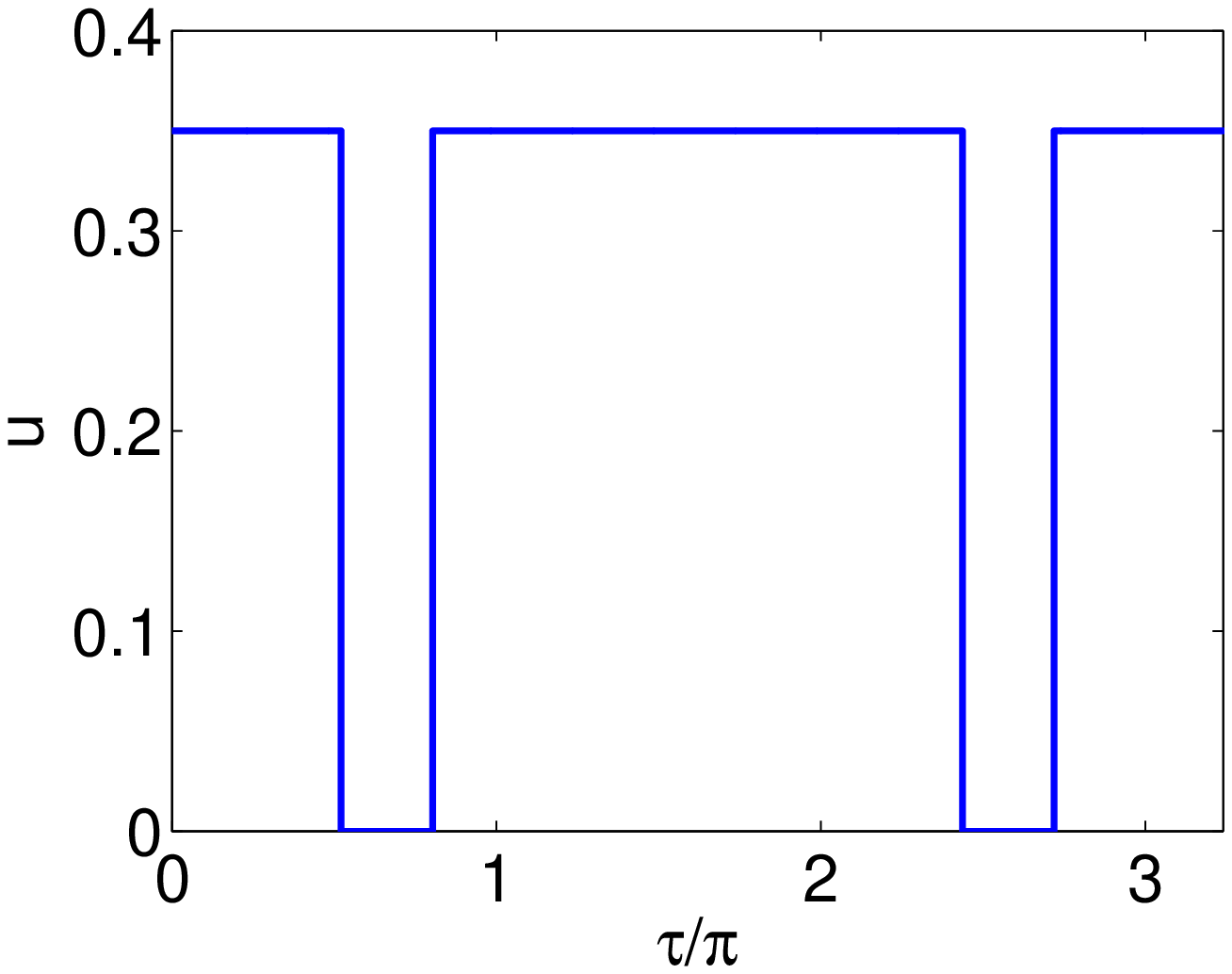}} \\
       \subfigure[$\ $]{
	            \label{fig:detuning}
	            \includegraphics[width=.45\linewidth]{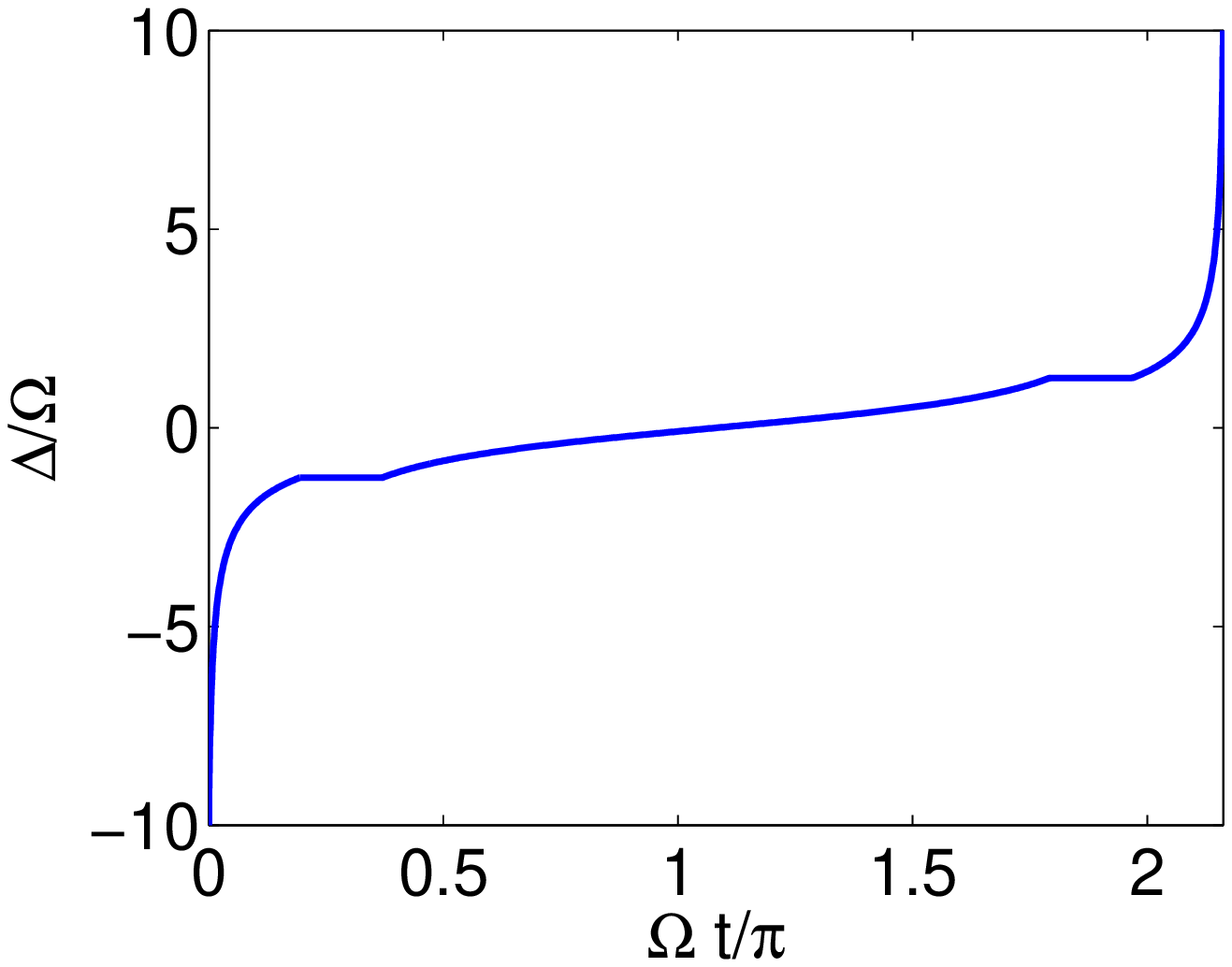}} &
       \subfigure[$\ $]{
	            \label{fig:angle}
	            \includegraphics[width=.45\linewidth]{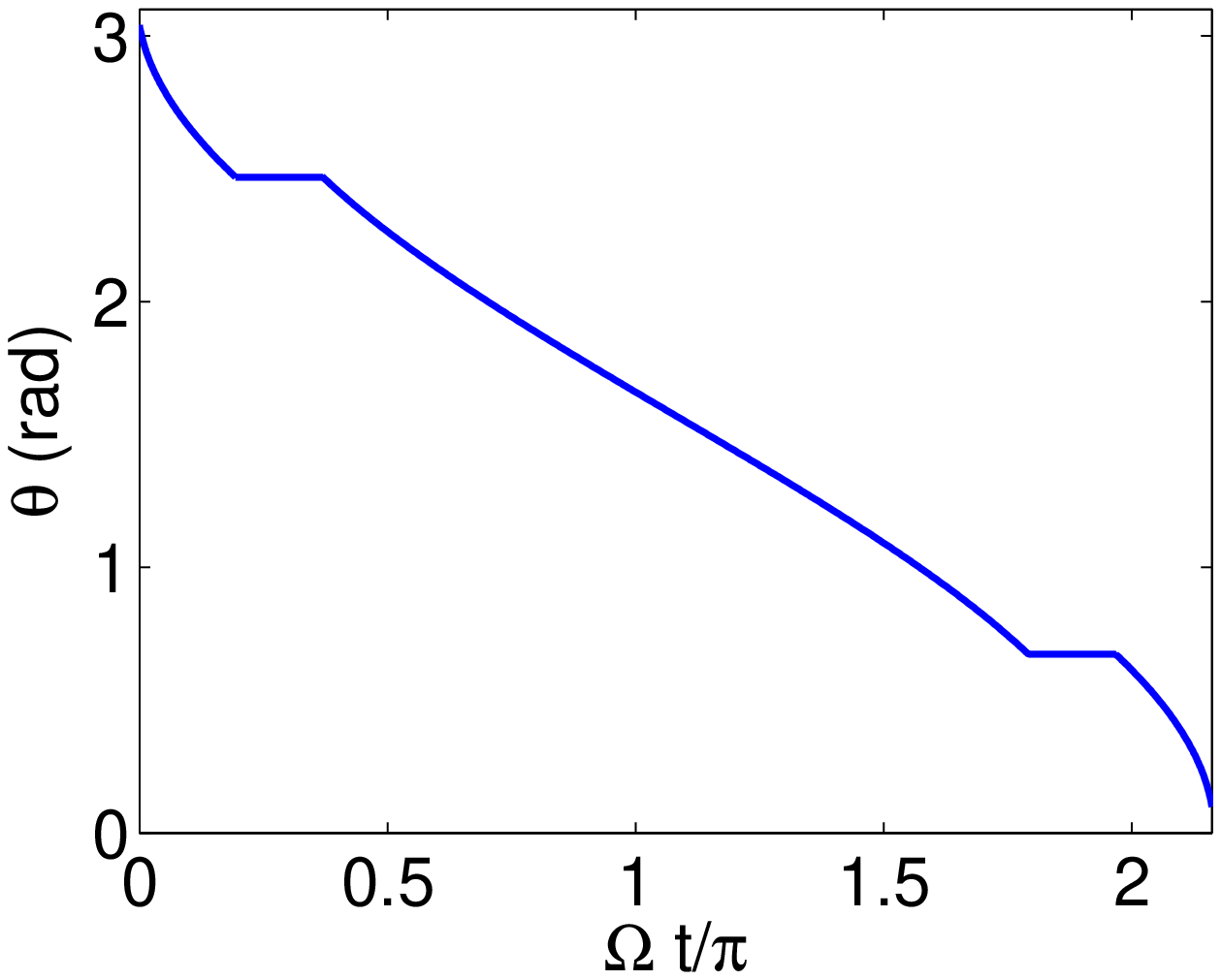}} \\
       \subfigure[$\ $]{
	            \label{fig:trajectory_original}
	            \includegraphics[width=.45\linewidth]{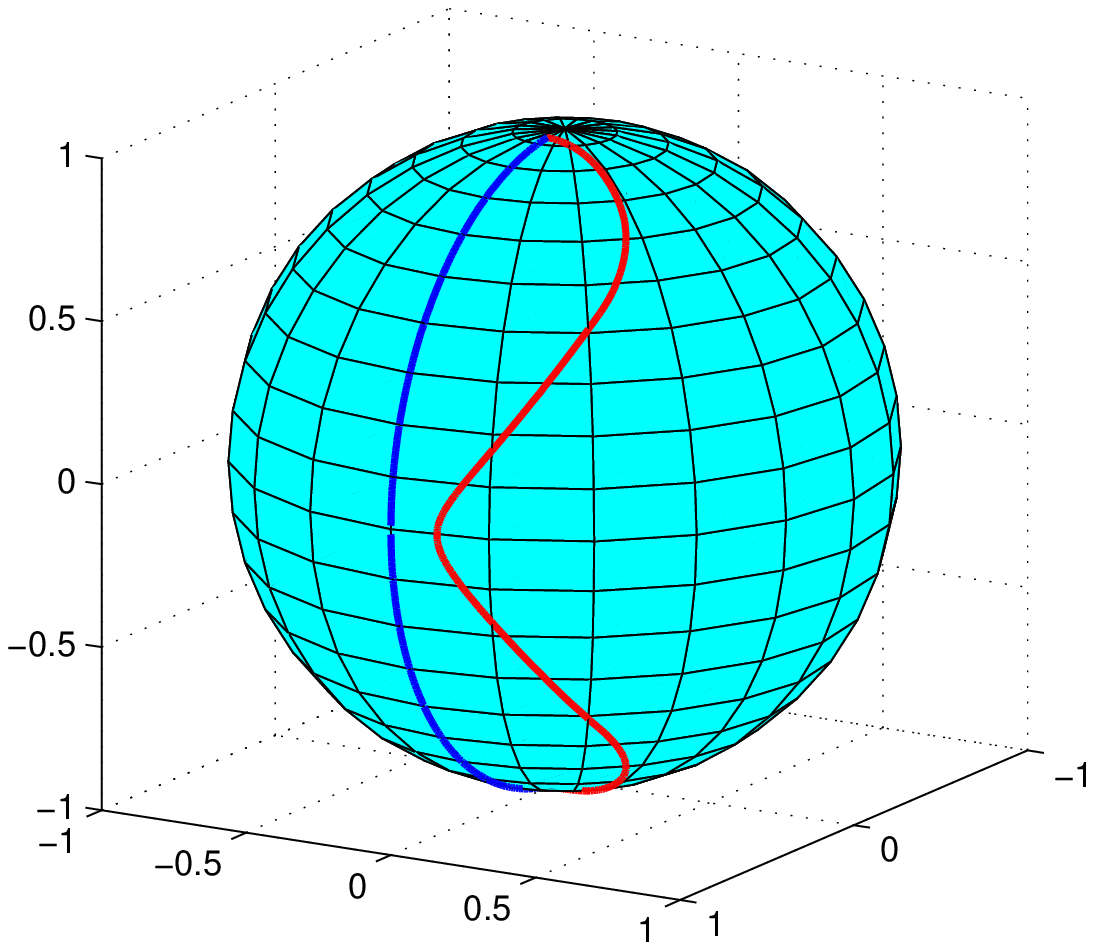}} &
       \subfigure[$\ $]{
	            \label{fig:trajectory_adiabatic}
	            \includegraphics[width=.45\linewidth]{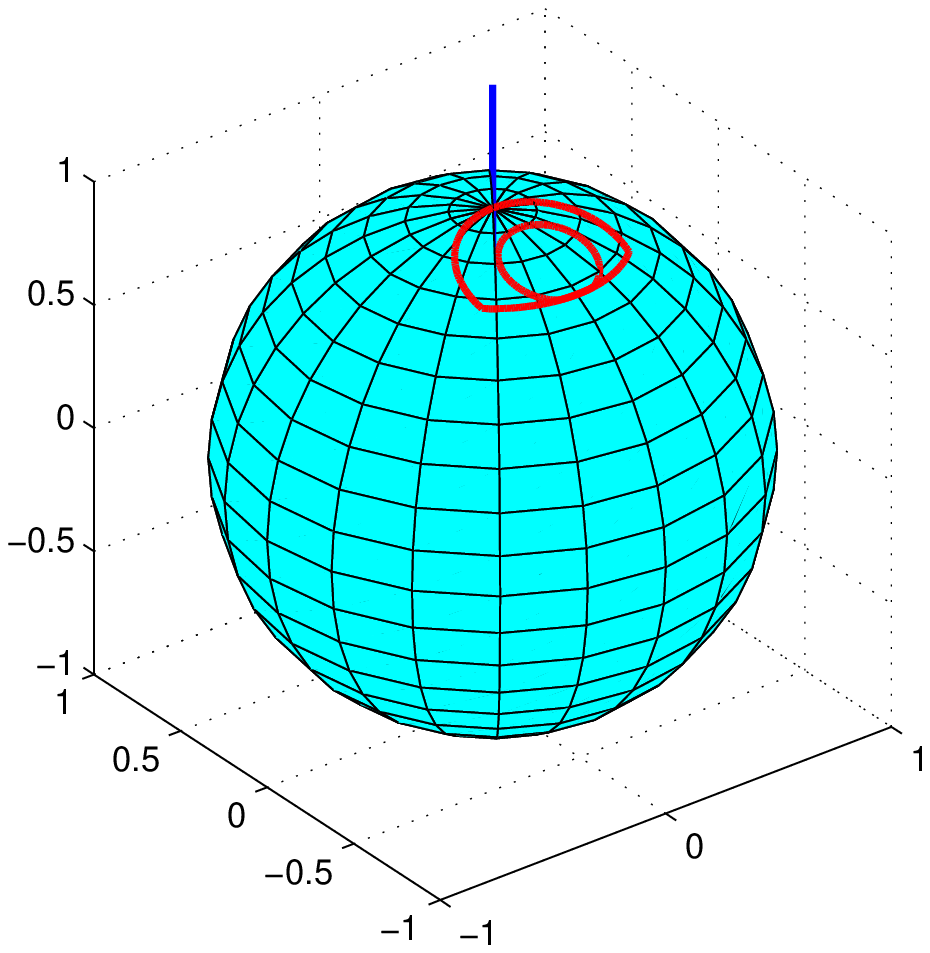}} \\
		\end{tabular}
\caption{Specific example for maximum control amplitude $v=0.35$, corresponding to the case highlighted with a red star in Fig. \ref{fig:durations}. (a) Logarithmic error $\log_{10}{(1-F)}=\log_{10}{|a_{y,2}|^2}$ as a function of duration $\tau_3$; the ``resonance" indicates the solution of the transcendental equation $a_{y,2}=0$. (b) Optimal pulse-sequence in the rescaled time $\tau$. (c) Detuning $\Delta(t)$ in the original time $t$. (d) Total field angle $\theta(t)$ in the original time $t$. (e) State trajectory (red solid line) on the Bloch sphere in the original reference frame. The blue solid line on the meridian lying on the $xz$-plane indicates the change in the total field angle $\theta$. (d) State trajectory (red solid line) on the Bloch sphere in the adiabatic frame. Observe that in this frame the state of the system returns to the north pole, while the total field points constantly in the $\hat{\bi{z}}$-direction (blue solid line).}
\label{fig:example}
\end{figure*}

As an example, we consider a change in the detuning from $\Delta_i=-10\Omega$ to $\Delta_f=10\Omega$, same as in \cite{Martinis14}, corresponding to $\theta_f=\tan^{-1}(1/10)$, $\theta_i=\pi-\theta_f$. In Fig. \ref{fig:durations} we plot the duration of the optimal pulse-sequence for a range of $v$ values, both in the rescaled time, Fig. \ref{fig:duration_rescaled}, and in the original time, Fig. \ref{fig:duration_original}. Note that the duration in the rescaled time is larger than the corresponding duration in the original time due to the sine factor in Eq. (\ref{rescaled_time}). The diagrams display a stairway-like form, where the circles separating the steps are the points $(u_k, T_k)$ obtained in Sec. \ref{sec:RC} where the original RC protocol, with constant control $u(\tau)=u_k$, is optimal. We have obtained similar diagrams in our other works on optimal control of quantum systems \cite{Stefanatos14,Stefanatos16}. On the first step from the right (larger values of $v$), the optimal pulse-sequence has the simple ``on-off-on" form, with $m=1$. Note that the solutions lying on this step are faster than the first resonance of the original RC protocol (first circle from the right). For large values of $v$ the duration of these solutions tends to the limit $T_0=\pi$. On the second step, the optimal pulse-sequence changes to ``on-off-on-off-on", with $m=2$. On the third step becomes ``on-off-on-off-on-off-on", with $m=3$, and so forth. Note that these solutions with more switchings may require longer times, but the corresponding maximum control amplitude $v$ is smaller and thus the change in the total field angle $\theta$ is less abrupt, a property which might be useful when designing a pulse-sequence.

In Fig. \ref{fig:example} we present a specific example of the optimal pulse-sequence for maximum control amplitude $v=0.35$, the case highlighted with a red star in Fig. \ref{fig:durations}. Since this point lies on the second step of the stairway-like diagram, the corresponding optimal pulse-sequence has the ``on-off-on-off-on" form. In Fig. \ref{fig:resonance} we display the logarithmic error
\begin{equation}
\label{logarithmic_error_1}
\log_{10}{(1-F)}=\log_{10}{|b_2(T)|^2}=\log_{10}{|a_{y,2}|^2}
\end{equation}
as a function of duration $\tau_3$; the ``resonance" indicates the solution of the transcendental equation $a_{y,2}=0$. Having found the duration $\tau_3$ of the intermediate ``on" pulse, we find the durations $\tau_1$ (of the initial and final ``on" pulses) and $\tau_2$ (of the ``off" pulses), using Eqs. (\ref{first_relation}), (\ref{second_relation}), respectively. In Fig. \ref{fig:sequence} we plot the optimal pulse-sequence $u(\tau)$ in the rescaled time $\tau$.
In Fig. \ref{fig:detuning} we show the detuning $\Delta(t)$ while in Fig. \ref{fig:angle} the corresponding evolution of the total field angle $\theta(t)$, both in the original time $t$. Note that the total duration in the rescaled time is larger than the corresponding duration in the original time due to the sine factor in Eq. (\ref{rescaled_time}). In Fig. \ref{fig:trajectory_original} we plot with red solid line the state trajectory on the Bloch sphere and in the original reference frame. The blue solid line on the meridian indicates the change in the total field angle $\theta$. Finally, in Fig. \ref{fig:trajectory_adiabatic} we plot the same trajectory (red solid line) but in the adiabatic frame. Note that in this frame the system starts from the adiabatic state at the north pole and returns there at the final time, while the total field points constantly in the $\hat{\bi{z}}$-direction (blue solid line). Also, observe that the trajectory in this frame contains a loop, which might look surprising at first sight for the solution of a minimum-time optimal control problem. The catch here is that there is actually an extra state variable not shown in this frame, the angle $\theta$, which evolves from $\theta_i$ to $\theta_f$. If the trajectory is displayed in the higher-dimensional space of all the state variables, the loop disappears.

We close this section by clarifying the advantage of the present approach compared to our previous related work \cite{Stefanatos19}. There, we fix the total duration $T=2\tau_1+m\tau_2+(m-1)\tau_3$ of the pulse-sequence in the rescaled time, while we take the amplitude $v$ as an unknown parameter. This relation, along with the pulse area condition (\ref{first_relation}) and the final condition (\ref{third_relation}), form a system of three equations with four unknowns, $\tau_1, \tau_2, \tau_3, v$. In order to tackle this problem, we find numerically the minimum value of the amplitude $v$ such that this system has a solution for $\tau_1, \tau_2, \tau_3$. In the present article we follow a dual approach, where we fix amplitude $v$ and seek the pulse-sequence with minimum duration which satisfies the area and final conditions. The use of optimal control theory leads to the optimality condition (\ref{optimality}) which, along with Eqs. (\ref{first_relation}) and (\ref{third_relation}), form a system of three equations for the three unknowns $\tau_1, \tau_2, \tau_3$.

\section{Conclusion}

\label{sec:conclusion}

In this article, we presented a new method for speeding up adiabatic passage in a two-level system with only detuning ($z$-field) control. This technique is actually a modification of the Roland-Cerf protocol, where now the local adiabaticity parameter is not held constant but has a simple ``on-off" modulation. Using optimal control theory, we found composite pulses which achieve perfect fidelity for every duration larger than the limit $\pi/\Omega$, where $\Omega$ is the constant transverse $x$-field. The corresponding detuning control is a continuous and monotonic function of time. The present work is expected to find applications in various tasks in quantum information processing, for example the design of high fidelity controlled-phase gates, but also in other research areas where adiabatic passage is exploited.

\section*{Acknowledgements}

The research is implemented through the Operational Program ``Human Resources Development, Education and Lifelong Learning'' and is co-financed by the European Union (European Social Fund) and Greek national funds (project E$\Delta$BM34, code MIS 5005825).

\section*{Appendix}

\label{appendix}

We first show that $a_x=0$ in Eq. (\ref{propagator}). From Eqs. (\ref{total_propagator}), (\ref{propagator}), and a well-known identity regarding the trace of a matrix product, we have
\begin{eqnarray}
\label{a_x}
a_x&=&\frac{1}{2}\mbox{Tr}(\sigma_xU)\nonumber\\
&=&\frac{1}{2}\mbox{Tr}(\sigma_xU_1W_2U_3\ldots W_2\,\mbox{or}\,U_3\ldots U_3W_2U_1)\nonumber\\
&=&\frac{1}{2}\mbox{Tr}(\ldots U_3W_2U_1\sigma_xU_1W_2U_3\ldots W_2\,\mbox{or}\,U_3).
\end{eqnarray}
But, using the explicit expressions (\ref{U}), (\ref{W}) for $U_1, W_2, U_3$ and the identity (\ref{Pauli}), it is not hard to verify that
\begin{equation}
U_1\sigma_xU_1=W_2\sigma_xW_2=U_3\sigma_xU_3=\sigma_x.
\end{equation}
Using the above relations repeatedly in Eq. (\ref{a_x}), it is not difficult to see that the calculation of $a_x$ is reduced to the calculation of $\mbox{Tr}(\sigma_xW_2)$ or $\mbox{Tr}(\sigma_xU_3)$, depending whether the middle pulse is ``off" or ``on", respectively. But $\mbox{Tr}(\sigma_xW_2)=\mbox{Tr}(\sigma_xU_3)=0$, thus $a_x=0$ as well.

We next explain how to find the coefficient $a_y$ in Eq. (\ref{propagator}). It is obtained from a relation similar to Eq. (\ref{a_x}),
\begin{eqnarray}
\label{a_y}
a_y&=&\frac{1}{2}\mbox{Tr}(\sigma_yU)\nonumber\\
&=&\frac{1}{2}\mbox{Tr}(\sigma_yU_1W_2U_3\ldots W_2\,\mbox{or}\,U_3\ldots U_3W_2U_1)\nonumber\\
&=&\frac{1}{2}\mbox{Tr}(\ldots U_3W_2U_1\sigma_yU_1W_2U_3\ldots W_2\,\mbox{or}\,U_3),
\end{eqnarray}
using repeatedly the equations
\begin{numparts}
\begin{eqnarray}
U_1\sigma_yU_1&=&in_y\sin{\omega \tau_1}I+(n_z^2+n_y^2\cos{\omega \tau_1})\sigma_y\nonumber\\
&&+n_yn_z(1-\cos{\omega \tau_1})\sigma_z,\\
W_2\sigma_yW_2&=&\sigma_y,\\
W_2\sigma_zW_2&=&-i\sin{\tau_2}+\cos{\tau_2}\sigma_z,\\
U_3\sigma_yU_3&=&in_y\sin{\omega \tau_3}I+(n_z^2+n_y^2\cos{\omega \tau_3})\sigma_y\nonumber\\
&&+n_yn_z(1-\cos{\omega \tau_3})\sigma_z,\\
U_3\sigma_zU_3&=&-in_z\sin{\omega \tau_3}I+n_yn_z(1-\cos{\omega \tau_3})\sigma_y\nonumber\\
&&+(n_y^2+n_z^2\cos{\omega \tau_3})\sigma_z,
\end{eqnarray}
\end{numparts}
which can be derived from expressions (\ref{U}) for $U_1,U_3$ and (\ref{W}) for $W_2$, as well as property (\ref{Pauli}).


\section*{References}

\begin{thebibliography}{99}


\bibitem{Roadmap17}
A. Ac\'{i}n {\it et al.} 2018 {\it New J. Phys.} {\bf 20} 080201

\bibitem{Glaser15}
Glaser S J {\it et al.} 2015 {\it Eur. Phys. J. D} {\bf 69} 279

\bibitem{Vitanov01}
Vitanov N V, Halfmann T, Shore B W and Bergmann K 2001 {\it Annu. Rev. Phys. Chem.} {\bf 52} 763

\bibitem{Goswami03}
Goswami D 2003 {\it Phys. Rep.} {\bf 374} 385

\bibitem{Martinis14a}
Kelly J {\it et al.} 2014 {\it Nature} {\bf 508} 500

\bibitem{Martinis14}
Martinis J M and Geller M R 2014 {\it Phys. Rev. A} {\bf 90} 022307

\bibitem{Shim16}
Shim Y P and Tahan C 2016 {\it Nat. Commun.} {\bf 7} 11059

\bibitem{Zeng18NJP}
Zeng J, Deng X H, Russo A and Barnes E 2018 {\it New J. Phys.} {\bf 20} 033011

\bibitem{Zeng18PRA}
Zeng J and Barnes E 2018 {\it Phys. Rev. A} {\bf 98} 012301

\bibitem{Fischer19}
Fischer J, Basilewitsch D, Koch C P and Sugny D 2019 {\it Phys. Rev. A} {\bf 99} 033410

\bibitem{Landau32}
Landau L 1932 {\it Phys. Z. Sowjetunion} {\bf 2} 46

\bibitem{Zener32}
Zener C 1932 {\it Proc. R. Soc. A} {\bf 137} 696

\bibitem{Garanin02}
Garanin D A and Schilling R 2002 {\it Phys. Rev. B} {\bf 66} 174438 

\bibitem{Calarco09}
Caneva T, Murphy M, Calarco T, Fazio R, Montangero S, Giovannetti V and Santoro G E 2009 {\it Phys. Rev. Lett.} {\bf 103} 240501

\bibitem{Hegerfeldt13}
Hegerfeldt G C 2013 {\it Phys. Rev. Lett.} {\bf 111} 260501

\bibitem{Hegerfeldt14}
Hegerfeldt G C 2014 {\it Phys. Rev. A} {\bf 90} 032110

\bibitem{Odelin19}
Gu\'{e}ry-Odelin D, Ruschhaupt A, Kiely A, Torrontegui E, Mart\'{i}nez-Garaot S and Muga J G 2019 Shortcuts to adiabaticity: concepts, methods, and applications arXiv:1904.08448

\bibitem{Demirplak03}
Demirplak M and Rice S A 2003 {\it J. Phys. Chem. A} {\bf 107} 9937

\bibitem{Berry09}
Berry M V 2009 {\it J. Phys. A: Math. Theor.} {\bf 42} 365303

\bibitem{Motzoi09}
Motzoi F, Gambetta J M, Rebentrost P and Wilhelm F K 2009 {\it Phys. Rev. Lett.} {\bf 103} 110501

\bibitem{Chen10a}
Chen X, Ruschhaupt A, Schmidt S, del Campo A, Gu\'{e}ry-Odelin D and Muga J G 2010 {\it Phys. Rev. Lett.} {\bf 104} 063002

\bibitem{Masuda10}
Masuda S and Nakamura K 2010 {\it Proc. R. Soc. A} {\bf 466} 1135

\bibitem{Deffner14}
Deffner S, Jarzynski C and del Campo A 2014 {\it Phys. Rev. X} {\bf 4} 021013

\bibitem{Chen10}
Chen X, Lizuain I, Ruschhaupt A, Gu\'{e}ry-Odelin D and Muga J G 2010 {\it Phys. Rev. Lett.} {\bf 105} 123003

\bibitem{Chen11}
Chen X, Torrontegui E and Muga J G 2011 {\it Phys. Rev. A} {\bf 83} 062116

\bibitem{Bason12}
Bason M G, Viteau M, Malossi N, Huillery P, Arimondo E, Ciampini D, Fazio R, Giovannetti V, Mannella R and Morsch O 2012 {\it Nat. Phys.} {\bf 8} 147

\bibitem{Malossi13}
Malossi N, Bason M G, Viteau M, Arimondo E, Mannella R, Morsch O and Ciampini D 2013 {\it Phys. Rev. A} {\bf 87} 012116

\bibitem{Ruschhaupt12}
Ruschhaupt A, Chen X, Alonso D and Muga J G 2012 {\it New J. Phys.} {\bf 14} 093040

\bibitem{Daems13}
Daems D, Ruschhaupt A, Sugny D and Gu\'{e}rin S 2013 {\it Phys. Rev. Lett.} {\bf 111} 050404 

\bibitem{Ibanez13}
Ib\'{a}\~{n}ez S, Chen X and Muga J G 2013 {\it Phys. Rev. A} {\bf 87} 043402

\bibitem{Motzoi13}
Motzoi F and Wilhelm F K 2013 {\it Phys. Rev. A} {\bf 88} 062318

\bibitem{Theis18}
Theis L S, Motzoi F, Machnes S and Wilhelm F K 2018 {\it EPL} {\bf 123} 60001

\bibitem{Roland02}
Roland J and Cerf N J 2002 {\it Phys. Rev. A} {\bf 65} 042308

\bibitem{Stefanatos19}
Stefanatos D and Paspalakis E 2019 Resonant shortcuts for adiabatic rapid passage with only $z$-field control arXiv:1906.11493 [quant-ph]

\bibitem{Levitt86}
Levitt M H 1986 {\it Prog. Nucl. Magn. Reson. Spectrosc.} {\bf 18} 61

\bibitem{Torosov11}
Torosov B T, Gu\'{e}rin S and Vitanov N V 2011 {Phys. Rev. Lett.} {\bf 106} 233001

\bibitem{Torosov18}
Torosov B T and Vitanov N V 2018 {\it Phys. Rev. A} {\bf 97} 043408

\bibitem{Torosov19}
Torosov B T and Vitanov N V 2019 {\it Phys. Rev. A} {\bf 99} 013402

\bibitem{Oh13}
Oh S, Shim Y P, Fei J, Friesen M, and Hu X 2013 {\it Phys. Rev. A} {\bf 87} 022332

\bibitem{Kral07}
Kr\'{a}l P, Thanopulos I and Shapiro M 2007 {\it Rev. Mod. Phys.} {\bf 79} 53

\bibitem{Vitanov17}
Vitanov N V, Rangelov A A, Shore B W and Bergmann K 2017 {\it Rev. Mod. Phys.} {\bf 89} 015006

\bibitem{Rezek09}
Rezek Y, Salamon P, Hoffmann K H and Kosloff R 2009 {\it EPL} {\bf 85} 30008

\bibitem{Kosloff17}
Kosloff R and Rezek Y 2017 {\it Entropy} {\bf 19} 136

\bibitem{Poggi13}
Poggi P M, Lombardo F C and Wisniacki D A 2013 {\it EPL} {\bf 104} 40005

\bibitem{Boscain02}
Boscain U, Charlot G, Gauthier J P, Gu\'{e}rin S and Jauslin H R 2002 {\it J. Math. Phys.} {\bf 43} 2107

\bibitem{Pontryagin}
Pontryagin L S, Boltyanskii V G, Gamkrelidze R V and Mishchenko E F 1962 {\it The Mathematical Theory of Optimal Processes} (New York: Interscience Publishers)

\bibitem{Schattler12}
Sch\"{a}ttler H and Ledzewicz U 2012 {\it Geometric Optimal Control: Theory, Methods and Examples} (Springer)

\bibitem{Sussmann87}
Sussmann H J 1987 {\it SIAM J. Control Optim.} {\bf 25} 433--65

\bibitem{Boscain05}
Boscain U and Chitour Y 2005 {\it SIAM J. Control Optim.} {\bf 44} 111--39

\bibitem{Boscain06}
Boscain U and Mason P 2006 {\it J. Math. Phys.} {\bf 47} 062101

\bibitem{Stefanatos14}
Stefanatos D and Li J S 2014 {\it IEEE Trans. Automat. Control} {\bf 59} 733--8

\bibitem{Stefanatos16}
Stefanatos D 2016 {\it Automatica} {\bf 73} 71--5













































\end{thebibliography}
\end{document}